\documentclass[preprint,aps,prl,showpacs,preprintnumbers, superscriptaddress, amsmath,amssymb]{revtex4}

\usepackage{stmaryrd} 
\usepackage{mathtools} 
\usepackage{stmaryrd} 
\usepackage{listings}
\usepackage[all]{xy}
\usepackage{amsmath}
\usepackage{amsfonts}
\usepackage{amssymb}
\usepackage{graphicx}%
\usepackage{bm}

\usepackage{listings}

\begin{document}

\newcommand{\nn}{\nonumber} 
\newcommand{\ms}[1]{\mbox{\scriptsize #1}}
\newcommand{\msi}[1]{\mbox{\scriptsize\textit{#1}}}
\newcommand{\dg}{^\dagger}
\newcommand{\smallfrac}[2]{\mbox{$\frac{#1}{#2}$}}
\newcommand{\ket}[1]{| {#1} \ra}
\newcommand{\bra}[1]{\la {#1} |}
\newcommand{\pfpx}[2]{\frac{\partial #1}{\partial #2}}
\newcommand{\dfdx}[2]{\frac{d #1}{d #2}}
\newcommand{\half}{\smallfrac{1}{2}}
\newcommand{\s}{{\mathcal S}}
\newcommand{\bs}[1]{\boldsymbol{#1}} 

\title{ Efficient method to generate time evolution of the Wigner function for open quantum systems}

\author{Renan Cabrera}
\email{rcabrera@princeton.edu}
\affiliation{Department of Chemistry, Princeton University, Princeton, NJ 08544, USA} 

\author{Denys I. Bondar}
\affiliation{Department of Chemistry, Princeton University, Princeton, NJ 08544, USA} 

\author{Kurt Jacobs}
\affiliation{ U.S. Army Research Laboratory, Computational and Information Sciences Directorate, Adelphi, Maryland 20783, USA}
\affiliation{ Hearne Institute for Theoretical Physics, Louisiana State University, Baton Rouge, Louisiana 70803, USA  }
\affiliation{ Department of Physics, University of Massachusetts at Boston, Boston, Massachusetts 02125, USA}

\author{Herschel A. Rabitz}
\affiliation{Department of Chemistry, Princeton University, Princeton, NJ 08544, USA} 

\date{\today}

\begin{abstract}
The Wigner function is a useful tool for exploring the transition between quantum and classical dynamics, 
as well as the behavior of quantum chaotic systems. Evolving the Wigner function for open systems 
has proved challenging however; a variety of methods have been devised but suffer from being 
cumbersome and resource intensive. Here we present an efficient fast-Fourier method for evolving the 
Wigner function, that has a complexity of $O(N\log N)$ where $N$ is the size of the array storing the Wigner function. The efficiency, 
stability, and simplicity of this method allows us to simulate open system dynamics previously thought to be prohibitively 
expensive. As a demonstration we simulate the dynamics of both one-particle and two-particle systems under various environmental 
interactions. For a single particle we also compare the resulting evolution with that of the classical Fokker-Planck and 
Koopman-von Neumann equations, and show that the environmental interactions induce the quantum-to-classical transition as 
expected. In the case of two interacting particles we show that an environment interacting with one of the particles 
leads to the loss of coherence of the other.  
\end{abstract}


\maketitle

\section{Introduction}
The Wigner function is a useful tool in understanding the relationship between quantum systems and their classical 
counterparts~\cite{zachos2005quantum, PhysRevD.58.025002, zachos2001, bolivar2004quantum, polkovnikov2010phase}, especially 
for chaotic systems in which visualization in phase-space has been crucial in enabling breakthroughs~\cite{Heller84}. The Wigner 
function is also very useful for studying the quantum-to-classical transition, the process in which classical dynamics emerges as 
an effective theory from the underlying quantum mechanics~\cite{Bhattacharya00, PhysRevLett.88.040402, Bhattacharya03, RevModPhys.75.715, Everitt09, jacobs2014quantum}, and for which open systems play an important role~\cite{Hillery1984121, Kapral2006, petruccione2002theory, Bolivar2012, caldeira2014introduction}. 

The equation of motion for the Wigner function is known as Moyal's equation, and can be written either as an infinite-order 
partial differential equation or as an integral equation~\cite{moyal1949quantum, groenewold1946principles}. Both forms 
are difficult to solve and, as a result, a plethora of numerical methods for evolving the Wigner function propagation 
have been developed. These have involved (i) the integral form of Moyal's equation~\cite{barker1983quasi, gronager1995quantum, wong2003explicit, dittrich2006semiclassical, dittrich2010semiclassical, sakurai2014self}, (ii) reduction of the Moyal equation to a Boltzmann-like equation~\cite{brosens2010newtonian, sels2012classical}, (iii) propagation of Gaussian and coherent 
states \cite{herman1984semiclasical, shimshovitz2012phase, dimler2009accurate, PhysRevLett.110.263202}, (iv) Monte Carlo 
schemes in which the Wigner function is contracted by averaging over stochastic trajectories of 
pure-states~\cite{shifren2001particle, PhysRevE.80.036701, filinov1995quantum, querlioz2006study}, and (v) 
evolving the density matrix in the coordinate representation~\cite{gao1997dissipative, grossmann2009finite}. 

In this paper we combine a recently developed, elegant formalism for quantum mechanics 
in phase space~\cite{bondar2012operational, PhysRevA.88.052108} with the spectral split operator method \cite{feit1982solution}. 
The spectral (fast Fourier transform) method is desirable because 
it allows one to take advantage of excellent existing libraries, parallelizes well, and is efficient 
and highly stable. The versatility and effectiveness of the resulting numerical method is illustrated 
by simulating decoherence and energy dissipation in single- and two-particle systems. 

The rest of the paper is organized as follows. The Hilbert phase space formalism that underlies the numerical methods is 
introduced in Sec.~\ref{MethodSec}. In this section we show how master equations for open systems are written in this 
formalism, as well as the evolution equations that describe classical motion. We also discuss the relationship between 
the equations describing the quantum and classical evolution. The split-operator technique for evolving the Wigner 
function is then presented in Sec. \ref{SplitOpSec}. In Secs. \ref{SingleParticleSim} and \ref{TwoParticleSim} we 
illustrate the use of the split operator technique by applying it to a number of examples. Section~\ref{conc} concludes with a brief summary. 

\section{ Formalism}
\label{MethodSec}

\subsection{Hilbert phase-space}

We first define the following notation. Given continuous variables $a$ and $b$, we write the derivatives 
with respect to these variables in the following compact form  
\begin{equation}
    \partial_a  \equiv \frac{\partial}{\partial a}, \;\;\; \partial_a^2  \equiv \frac{\partial^2}{\partial a^2}, \;\;\; \partial_{ab}^2  \equiv \frac{\partial^2}{\partial a \partial b}
\end{equation}
We will also use $\bs{a}$ to denote a continuous variable that is distinct from $a$. As is common we use hats to denote 
quantum operators that correspond to classical observables. Thus, the position operator $\hat{\bs{x}}$ has a continuous spectrum of 
eigenvalues given by the variable $\bs{x}$, and the corresponding momentum operator is $\hat{\bs{p}} \equiv -i \hbar \bs{\partial}_{\bs{x}}$. We do not use a hat for 
the density operator, which we denote by $\rho$, and we write the matrix elements of $\rho$ in the compact 
form $\rho_{\bs{xy}} = \langle \bs{x} | \rho | \bs{y}\rangle $. Finally, for a function $f$ of two variables $x$ and $y$, we use the form $f(x,y)$ as well as 
the more compact form $f_{xy}$.  

With the above notation the unitary evolution for the quantum density operator $\rho$ is given by~\cite{gardiner2004quantum}
\begin{align}\label{RhoEq}
 i\hbar \dot{\rho} = [ \hat{H}( \hat{\boldsymbol x} , \hat{ \boldsymbol p} ) , \rho ].
\end{align}
where $[\hat{\boldsymbol x},\hat{\boldsymbol p}] = i\hbar$ and $\hat{H}$ is the Hamiltonian. In particular, Eq. (\ref{RhoEq}) 
in the position representation is
\begin{align}
\label{RhoEq2}
 i \hbar \, \partial_t \rho_{\bs{xx'}} 
 =  \left[ H \left( \bs{x} , -i \hbar \bs{\partial}_{\bs{x}} \right)  
  -  H \left( \bs{x}^\prime  , i \hbar \bs{\partial}_{\bs{x'} } \right) 
   \right] \rho_{\bs{xx'}} . 
\end{align}
The linear change of variables, 
\begin{align}
  \bs{x} =  x -\smallfrac{\hbar}{2} \theta, \qquad
  \bs{x}^\prime =  x +\smallfrac{\hbar}{2} \theta,
\end{align}
gives the new representation 
\begin{align}
	B_{x\theta} = \langle  x - \smallfrac{\hbar}{2}\theta | \rho  | x + \smallfrac{\hbar}{2}\theta  \rangle, 
	\label{B-def}  
\end{align}
with the new equation of motion 
\begin{align}
 i \hbar \, \partial_t  B_{x\theta} 
 = & \left[ 
    H\left( x - \smallfrac{\hbar}{2}\theta  , i \left[ \partial_\theta  - \smallfrac{\hbar}{2} \partial_x \right] \right)  \right.  
   \left. - H\left( x + \smallfrac{\hbar}{2}\theta , i \left[ \partial_\theta  + \smallfrac{\hbar}{2} \partial_x \right] \right) 
   \right] B_{x\theta} .    \label{B-eq}
\end{align}
Since $ \hbar \theta$ has dimensions of length, the function $B_{x\theta}$ was named the ``double configuration space representation'' 
by Blokhintsev~\cite{Blokhintsev1940a,Blokhintsev1941}. Following Blokhintsev it is possible to define
the quantity $p$, with dimensions of momentum, as the conjugate variable to $\theta$.  
In this way we obtain the celebrated Wigner function, $W_{xp}$, 
related to $B_{x\theta}$ through the Fourier transform:   
\begin{align}
 \label{W-to-B}
 B_{x\theta} = & \int    W_{xp} \, e^{-ip\theta} dp , \qquad \\ 
  W_{xp} = & \frac{1}{2\pi} \int   B_{x\theta} \, e^{i p \theta}  d \theta  \label{B-to-W} . 
\end{align}
Note that while $B_{x\theta}$ is in general a complex valued function, $W_{xp}$ is real 
and can be normalized according to
\begin{align}
\label{W-normalization}
  \int W_{xp} \, dx \, dp = 1.
\end{align}
Nevertheless, considering that $W_{xp}$ is not necessarily positive, it cannot be interpreted 
as a true probability distribution (see discussions below).

Using the above definitions we obtain the equation of motion in phase space 
\begin{align}
\label{W-eq}
 i \hbar \, \partial_t  W_{xp} 
 =  & {[} 
    H\left( x + \smallfrac{i\hbar}{2}\partial_p  , p  - \smallfrac{i\hbar}{2} \partial_x \right)  
  - H\left( x - \smallfrac{i\hbar}{2}\partial_p , p  + \smallfrac{i\hbar}{2} \partial_x \right) 
   {]} W_{xp}. 
\end{align}
The latter can be also expressed in terms of the Moyal star defined as
\begin{align}
  H_{xp} \star W_{xp} \equiv H_{xp} \exp \! \left( 
  \smallfrac{i\hbar}{2} \, \overleftarrow{\partial_x} \, 
  \overrightarrow{\partial_p} - 
   \smallfrac{i\hbar}{2} \, 
   \overleftarrow{\partial_p} \, \overrightarrow{\partial_x}
  \right)  W_{xp}, 
\end{align}
where the arrows indicate the direction of the derivatives' action, and we have
 written $H_{xp} \equiv H(x,p)$. Employing the following identities 
\begin{align}
H_{xp} \star W_{xp} &= H\!\left(x + \smallfrac{i\hbar}{2} \overrightarrow{\partial_p} 
, p - \smallfrac{i\hbar}{2} \overrightarrow{\partial_x} \right) W_{xp}, \nonumber \\
W_{xp} \star H_{xp}  &= H\!\left(x - \smallfrac{i\hbar}{2}\overrightarrow{\partial_p}, 
p + \smallfrac{i\hbar}{2} \overrightarrow{\partial_x} \right) W_{xp},
\end{align}
the equation of motion (\ref{W-eq}) becomes
\begin{align}
 i \hbar \, \partial_t W_{xp} = H_{xp} \star W_{xp} - W_{xp} \star H_{xp}, 
\end{align}
which is Moyal's equation~\cite{moyal1949quantum, zachos2005quantum, curtright2012quantum}.

An abstract formalism that is independent of the particular representation  can be introduced by
defining an extended four-operator algebra  $\hat{x},\hat{p}, \hat{\theta}, \hat{\lambda}$ satisfying the following commutator
 relations  \cite{bondar2012operational, PhysRevA.88.052108}:  
\begin{align}
\label{commutation-rels}
 {[} \hat{x} , \hat{p} {]} = 0,  \quad
 {[} \hat{x} , \hat{\lambda} {]} = i, \quad
 {[} \hat{p} , \hat{\theta} {]} = i, \quad
 {[} \hat{\lambda} , \hat{\theta} {]} = 0.
  \end{align}
We note that the commuting operators $\hat{x}$ and $\hat{p}$, representing position and momentum in the phase space, form a basis 
for the Koopman-von Neumann representation of classical mechanics \cite{Koopman1931, Neumann1932, Neumann1932a, DaniloMauro2002}. 
The operators $\hat{\lambda}$ and $ \hat{\theta}$ are known as the Bopp operators \cite{Bopp1956,Hillery1984121}. 
The four operators (\ref{commutation-rels}) can be used to realize the usual canonically-conjugate 
position and momentum coordinates via 
\begin{align}
 \label{xp-operators}
   \hat{\bs{x}} = \hat{x} - \smallfrac{\hbar}{2} \hat{\theta} ,\qquad
   \hat{\bs{p}} = \hat{p} + \smallfrac{\hbar}{2} \hat{\lambda}, 
\end{align}
so that $[ \hat{\bs{x}}  , \hat{\bs{p}}  ]  =  i \hbar $. Similarly, one can define a mirror quantum algebra as 
\begin{align}
 \label{xp-operators-mirrow}
   \hat{\bs{x}}^\prime  = \hat{x} + \smallfrac{\hbar}{2} \hat{\theta} ,\qquad
   \hat{\bs{p}}^\prime  = \hat{p} - \smallfrac{\hbar}{2} \hat{\lambda}, 
\end{align}
obeying the commutation relation with the negative sign $[ \hat{\bs{x}}^\prime , \hat{\bs{p}}^\prime ]  =  -i \hbar $, while all the other 
commutators vanish: $  [ \hat{\bs{x}} ,  \hat{\bs{x}}^\prime  ]
   =  [ \hat{\bs{x}} ,  \hat{\bs{p}}^\prime  ] 
   =  [ \hat{\bs{p}}^\prime  ,  \hat{\bs{p}}  ] 
   =  [ \hat{\bs{p}}^\prime  ,  \hat{\bs{x}}  ] = 0  $.

The four operators $\hat{x}$, $\hat{\theta}$, $\hat{\lambda}$, and $\hat{p}$ can be used to define a Hilbert space that we refer 
to as the ``Hilbert phase space'' after~\cite{PhysRevA.88.052108}. Specifically, since the self-adjoint 
operators $\hat{x}$ and $\hat{\theta}$ (respectively $\hat{\lambda}$ and $\hat{p}$) commute, they share a common orthonormal 
eigenbasis $| x \theta \rangle $ (respectively $| \lambda p \rangle $). These bases are complete so naturally 
\begin{align}
 \bs{1} = \int dx d\theta    \scalebox{1.3}{\ensuremath{ \bm |}}   x \theta   \scalebox{1.4}{\ensuremath{ \bm \rangle }}
     \scalebox{1.3}{\ensuremath{\bm \langle }} x \theta  
 \scalebox{1.4}{\ensuremath{ \bm | }} = \int d\lambda dp   \scalebox{1.3}{\ensuremath{ \bm |}}   \lambda p 
 \scalebox{1.4}{\ensuremath{ \bm \rangle }}
  \scalebox{1.3}{\ensuremath{\bm \langle }} \lambda p  
  \scalebox{1.4}{\ensuremath{ \bm | }}, 
\end{align}
where $   \scalebox{1.3}{\ensuremath{\bm \langle }}  \lambda p \scalebox{1.3}{\ensuremath{\bm | }}
 x \theta  \scalebox{1.3}{\ensuremath{\bm \rangle }} = \exp( i p \theta - i x \lambda  )/(2\pi)  $. 

The position and momentum coordinates introduced above, as well as their mirror counterparts allow Eq.(\ref{RhoEq2}) to be rewritten in the more abstract form 
\begin{align}
\label{Coherent-evolution}
 i \hbar \frac{d}{ d t}  
  \scalebox{1.3}{\ensuremath{ \bm |}} 
  \rho 
  \scalebox{1.4}{\ensuremath{ \bm \rangle }} 
 =  \left[ H\left( \hat{\boldsymbol x} ,  \hat{\bs{p}} \right)  
  - H\left( \hat{ \bs{x} }^\prime   ,  \hat{\bs{p}}^\prime \right) 
   \right]    \scalebox{1.3}{\ensuremath{ \bm |}}   \rho   \scalebox{1.3}{\ensuremath{\bm \rangle }}  , 
\end{align}
where $  \scalebox{1.2}{\ensuremath{ \bm |}}   \rho   \scalebox{1.3}{\ensuremath{ \bm \rangle }}  $ is a ket belonging to the Hilbert phase space . 

We can realize  $\hat{x}, \hat{p}, \hat{\theta}$, and $\hat{\lambda}$ in terms of differential operators. 
For example, the phase space representation $x-p$ is accomplished by 
\begin{align}
 \label{phase-space-rep}
 \hat{x} = x, \quad
 \hat{p} = p, \quad
 \hat{\lambda} = -i {\partial_x}, \quad
 \hat{\theta} = -i { \partial_p}, 
\end{align}
while, the $x-\theta$ representation requires
\begin{align}
 \hat{x} = x, \quad
 \hat{p} = i {\partial_\theta}, \quad
 \hat{\lambda} = -i {\partial _x}, \quad
 \hat{\theta} = \theta.  
\end{align}
Other representations can be constructed in a similar fashion.

The Hilbert phase space formalism conveniently unites previously known results regarding phase-space 
distribution functions. Considering the Hamiltonian form $\hat{H} = \frac{1}{2m}\hat{\boldsymbol p}^2 + V(\hat{ \boldsymbol x})$, the abstract equation of motion 
for the density matrix is 
\begin{align}
\label{gen-motion}
 i \hbar \frac{d }{d t}
 \scalebox{1.3}{\ensuremath{ \bm |}} \rho   \scalebox{1.4}{\ensuremath{ \bm \rangle }}  
  = 
   \left[ \frac{\hbar}{m} \hat{p} \hat{\lambda} + 
   V\left(\hat{x} - \smallfrac{\hbar}{2} \hat{\theta } \right) 
-  V\left( \hat{x} + \smallfrac{\hbar}{2} \hat{\theta} \right) \right] 
  \scalebox{1.3}{\ensuremath{ \bm |}} \rho \scalebox{1.4}{\ensuremath{ \bm \rangle }} 
,
\end{align}
for which the $x-\theta$ representation gives a linear partial differential equation
\begin{align}
\label{B-equation-D}
i\hbar \,\partial_t \scalebox{1.4}{\ensuremath{ \bm | }}  \rho 
 \scalebox{1.4}{\ensuremath{ \bm \rangle }}_{\!\! x \theta}
 =  \left[ \frac{\hbar}{m}\partial^2_{x\theta} 
 + V \left( x - \smallfrac{\hbar}{2}\theta  \right) - V\left( x + \smallfrac{\hbar}{2} \theta  \right)
\right] \scalebox{1.4}{\ensuremath{ \bm | }}  \rho \scalebox{1.4}{\ensuremath{ \bm \rangle }}_{\!\! x \theta},
\end{align}
where $\scalebox{1.4}{\ensuremath{ \bm | }}  \rho \scalebox{1.4}{\ensuremath{ \bm \rangle }}_{\!\! x \theta} \equiv \scalebox{1.4}{\ensuremath{ \bm \langle }} x \theta    \scalebox{1.4}{\ensuremath{ \bm | }}  \rho 
 \scalebox{1.4}{\ensuremath{ \bm \rangle }}$. Since this differential equation is the same as 
Eq.(\ref{B-eq}) we have~\cite{PhysRevA.88.052108}
 \begin{equation}
   B(x,\theta) = \frac{1}{\sqrt{\hbar}} \scalebox{1.4}{\ensuremath{ \bm | }}  \rho 
 \scalebox{1.4}{\ensuremath{ \bm \rangle }}_{\!\! x \theta}  . 
\end{equation}
Alternatively, the same equation in the usual phase space is
\begin{align}
\label{Moyal-Eq} 
 i \hbar {\partial_t} \scalebox{1.4}{\ensuremath{ \bm | }}  \rho \scalebox{1.4}{\ensuremath{ \bm \rangle }}_{\!\! xp}  = 
   \biggl[ -i \frac{\hbar}{m} p {\partial_x}  & + V^+ -  V^-  \biggr]  
 \scalebox{1.4}{\ensuremath{ \bm | }}  \rho \scalebox{1.4}{\ensuremath{ \bm \rangle }}_{\!\! xp},
\end{align}
where $V^\pm = V\left(x \pm i \smallfrac{\hbar}{2} { \partial _p } \right)$ and 
\begin{equation}
 W(x,p) = \frac{1}{\sqrt{2 \pi \hbar}} \scalebox{1.4}{\ensuremath{ \bm | }}  \rho \scalebox{1.4}{\ensuremath{ \bm \rangle }}_{\!\! xp} .
\end{equation}

Equations (\ref{B-equation-D}) and (\ref{Moyal-Eq}) illustrate the power of 
choosing an appropriate representation: The equation of motion in the $x-\theta$ representation is a 
second-order partial differential equation with the same computational complexity as the
two-dimensional Schr\"odinger equation, while the equation of motion in the $x$-$p$ representation
is much more difficult to solve, as either a higher order partial differential equation  or
an equally challenging integro-differential equation \cite{zachos2005quantum}. 

In addition to $W(x,p)$ [$x$-$p$ phase space] and $B(x,\theta)$ [$x$-$\theta$ space], the quantum 
state can be represented by the functions $A(\lambda,\theta)$ and $Z(\lambda,p)$  
as
\begin{align}
 A(\lambda,\theta) &= \int dx \, e^{-i \lambda x}  B(x,\theta), \\
 Z(\lambda,p) & =  \frac{1}{2\pi} \int dx d\theta \, e^{i ( p\theta - \lambda x)} B(x,\theta),
\label{Z-equation}
\end{align}
where $A(\lambda,\theta)$ is known as the ambiguity function in signal-processing~\cite{cohen1985generalized},
and $Z(\lambda,p)$ can be regarded as the double momentum space representation since $\hbar\lambda$ has the dimensionality 
of momentum. The connections among all these functions are easily visualized in the following diagram:
{\large
\begin{align}
  \label{Fourier-transforms}
    \xymatrix{
      W(x,p)   &   Z(\lambda,p)  \ar[l]_{ \mathcal{F}^{ \lambda \rightarrow x  } }  \\
      B(x,\theta)  \ar[u]^{\mathcal{F}_{  \theta \rightarrow  p   }}  &   A(\lambda,\theta)  \ar[u]_{ \mathcal{F}_{  \theta \rightarrow  p   }}  \ar[l]_{ \mathcal{F}^{ \lambda \rightarrow x  } }   }
\end{align}} 
where vertical arrows denote the $\theta \to p$ partial Fourier transforms (
${\mathcal{F}_{  \theta \rightarrow  p     }} $), while horizontal arrows indicate  the $\lambda \to x$ partial Fourier transforms( $ \mathcal{F}^{ \lambda \rightarrow x  } $).

\subsection{Open systems}

Having reviewed the equations of motion for unitary evolution in the Hilbert phase-space formalism, we now show 
how to write various standard Markovian master equations in this formalism. If a master equation that describes 
the interaction with an environment is time-independent then to preserve the positivity of the density matrix it must 
have the Lindblad form. This means that in addition to the unitary evolution the derivative of $\rho$ contains one or 
more additional terms of the form~\cite{gardiner2004quantum} 
\begin{align}
   \mathcal{L}[\rho ]  =   A \rho A^\dagger -\smallfrac{1}{2}  A^\dagger A \rho - \smallfrac{1}{2} \rho A^\dagger A ,      
\label{abstract-Lindblad}                                             
\end{align}
where $A$ can be any operator. For a single particle every operator $A$ can be written as a function of the position 
and momentum operators, so we can write $A(\hat{\bs{ x}},\hat{\bs{ p}})$. Following the steps leading to Eq.(\ref{Coherent-evolution}), 
each of the terms in the  Lindblad form $\mathcal{L}[\rho]$ can be easily translated to the Hilbert phase-space formalism by 
using the following rules:  
\begin{align}
  A(\hat{\bs{ x}},\hat{\bs{ p}}) \rho  \; & \Leftrightarrow \;  A( \hat{\bs{x}} , \hat{\boldsymbol p})  \scalebox{1.3}{\ensuremath{ \bm |}}  \rho \scalebox{1.4}{\ensuremath{ \bm \rangle }}   \\
 \rho A(\hat{\bs{ x}},\hat{\bs{ p}})  \; & \Leftrightarrow \;  A( \hat{\bs{ x}}^\prime ,\hat{ \boldsymbol p}^\prime )  \scalebox{1.3}{\ensuremath{ \bm |}}  \rho \scalebox{1.4}{\ensuremath{ \bm \rangle }} ,  
\end{align}
and the fact that $A( \hat{\bs{x}} , \hat{\boldsymbol p})$ commutes with $B( \hat{\bs{x}}^\prime, \hat{ \boldsymbol p}^\prime )$ for every $A$ and $B$. That is, when any 
operator $A(\hat{\bs{ x}},\hat{\bs{ p}})$ acts to the right on $\rho$ it acts on $\scalebox{1.3}{\ensuremath{ \bm |}}  \rho \scalebox{1.4}{\ensuremath{ \bm \rangle }}$ as itself, and when it acts to 
the left on $\rho$ it acts on $\scalebox{1.3}{\ensuremath{ \bm |}}  \rho \scalebox{1.4}{\ensuremath{ \bm \rangle }}$ as $A( \hat{\bs{ x}}^\prime ,\hat{ \boldsymbol p}^\prime )$. Note also that in the Hilbert phase-space  
\begin{align}
A( \hat{\bs{x}} , \hat{\boldsymbol p}) & =  A \left( \hat{x} - \frac{\hbar\hat{\theta}}{2}, \hat{p} + \frac{\hbar \hat{\lambda}}{2} \right) , \\
 A( \hat{\bs{ x}}^\prime ,\hat{ \boldsymbol p}^\prime )&=  A \left(\hat{x} + \frac{\hbar\hat{\theta}}{2}, \hat{p} - \frac{\hbar \hat{\lambda}}{2}  \right) . 
   \label{abstract-Lindblad2}
\end{align}  
As an example, the Lindblad operator for the Wigner function is 
\begin{widetext}
\begin{align} 
  \mathcal{L}[ W_{x,p} ]   = &
  \biggl[ A \left( x - \smallfrac{\hbar}{2} \hat{\theta},  p + \smallfrac{\hbar }{2} \hat{\lambda}\right) 
   A^\dagger \left( x + \smallfrac{\hbar}{2} \hat{\theta}, p - \smallfrac{\hbar }{2} \hat{\lambda} \right)  -\smallfrac{1}{2} A^\dagger \left( x - \smallfrac{\hbar\hat{\theta}}{2}, p + \smallfrac{\hbar \hat{\lambda} }{2} \right) 
   A \left( x - \smallfrac{\hbar\hat{\theta}}{2}, p + \smallfrac{\hbar \hat{\lambda}}{2}  \right)  \biggr. \nonumber \\
 &- \smallfrac{1}{2} A^\dagger \left( x + \smallfrac{\hbar\hat{\theta}}{2},  p - \smallfrac{\hbar \hat{\lambda} }{2} \right) 
   A \left( x + \smallfrac{\hbar\hat{\theta}}{2},  p - \smallfrac{\hbar \hat{\lambda}}{2}  \right)  \biggl. \biggr] W_{x,p} ,  
   \label{Wigner-Lindblad}
\end{align}  
\end{widetext}
where $ \hat{\theta} = -i { \partial_p } $ and $ \hat{\lambda} = -i { \partial_x } $.

We now give useful forms for two important master equations. The first is decoherence in the 
basis of $x$ for which the master equation is~\cite{RevModPhys.75.715, PhysRevLett.88.040402, gardiner2004quantum}
\begin{align}
i \hbar \mathcal{L}[\rho] &= - \frac{D}{\hbar^2} [\hat{\bs{x}} , [ \hat{\bs{x}} ,\rho] ]  =
  \frac{2D}{\hbar^2} \left( \hat{\boldsymbol x}  \rho \hat{\boldsymbol x} 
- \frac{1}{2} \hat{\boldsymbol x}^2  \rho
- \frac{1}{2} \rho \, \hat{\boldsymbol x}^2  \right);
\end{align}
however, it has a particularly simple form in the Hilbert phase space
\begin{align}
i \hbar \mathcal{L} \left[  
 \scalebox{1.4}{\ensuremath{ \bm | }} \rho  \scalebox{1.4}{\ensuremath{ \bm \rangle }} 
  \right] 
 =& \frac{D}{\hbar^2} \left[ 2(\hat{x}- \hbar\hat{\theta}/2) (\hat{x}+ \hbar\hat{\theta}/2) 
      \right. \notag  \\
   &  \,\,\,\, - (\hat{x}- \hbar\hat{\theta}/2) (\hat{x}- \hbar\hat{\theta}/2 )  \notag\\ 
   &  \,\,\,\, - \left. (\hat{x} +\hbar\hat{\theta}/2 ) (\hat{x} + \hbar\hat{\theta}/2 ) \right] 
 \scalebox{1.4}{\ensuremath{ \bm | }} \rho  \scalebox{1.4}{\ensuremath{ \bm \rangle }} 
 \notag\\
   = & - D \hat{\theta}^2  \scalebox{1.4}{\ensuremath{ \bm | }} \rho  \scalebox{1.4}{\ensuremath{ \bm \rangle }}  . 
   \label{ExtendedLindbladDecoherence}
\end{align}
As a result, Blokhintsev's dynamical equation for a quantum system undergoing decoherence in the position basis reads
\begin{align}
\label{B-D-eq}
 \partial_t B_{x\theta}  = &  \left[ \frac{-i}{m}\partial^2_{\theta x } + \frac{V^{-} - V^{+}}{i\hbar} - D \theta^2
\right] B_{x\theta}, 
\end{align}
with $V^{-} = V( x - \hbar \theta/2 )$ and  $V^{+} = V( x + \hbar \theta/2 )$. 

Another widely used master equation is the time-independent approximation to the Caldeira-Legget 
model~\cite{Dekker1977,Caldeira1983,gardiner2004quantum,Bolivar2012}. This master equation is not 
strictly correct because it is not in the Lindblad form, but it is good enough for many purposes to describe damping 
and thermalization of a harmonic oscillator~\cite{Munro1996}. It is given by  
\begin{equation}
\label{D-Caldeira-Legget}
 i\hbar \hat{\mathcal{D}}[\rho] = 
- \frac{i\gamma}{\hbar} [\bs{x}, [\bs{p}, \rho  ]_{+} ] 
- \frac{2m \gamma k T}{ \hbar^2  } [\bs{x}, [\bs{x} , \rho] ],
\end{equation}
Here $[\bs{p}, \rho  ]_{+}$  denotes the anticommutator, $\gamma$ is the damping 
coefficient and $T$ is the temperature of a bath. The Hilbert phase-space form of this 
master equation is 
\begin{align}
\label{B-Lindbladian-Caldeira-Legget-abstract}
i\hbar \hat{\mathcal{D}}
  \scalebox{1.4}{\ensuremath{ \bm | }} \rho  \scalebox{1.4}{\ensuremath{ \bm \rangle }}
= 2 \gamma ( i \hat{\theta} \hat{p} - m  kT \hat{\theta}^2  )
  \scalebox{1.4}{\ensuremath{ \bm | }} \rho  \scalebox{1.4}{\ensuremath{ \bm \rangle }} ,
\end{align}
and in the $x$-$\theta$ representation this becomes 
\begin{align}
\label{Caldeira-Legget-eq}
\partial_t B_{x\theta}
 =&  \left[
\frac{-i}{m} \partial^2_{x\theta}  
 + \frac{V^{-} - V^{+}}{i\hbar} - 2 \gamma \theta \left(
    \partial_\theta + m k T \theta 
\right)  
\right] B_{x\theta} .
\end{align}


\subsection{Hilbert Phase Space and Classical Dynamics}
Classical mechanics can be embedded in the Hilbert phase space. As discussed 
in Ref.~\cite{PhysRevA.88.052108}, when we take the classical limit $\hbar \to 0$ of Eq. (\ref{gen-motion}) we recover the 
Koopman-von Neumann equation for the classical state $ \scalebox{1.4}{\ensuremath{ \bm | }}  \rho 
 \scalebox{1.4}{\ensuremath{ \bm \rangle }}$~\cite{Koopman1931, Neumann1932, Neumann1932a, DaniloMauro2002}
\begin{align}
\label{Koopman-Eq}
 i \frac{d }{d t}
 \scalebox{1.3}{\ensuremath{ \bm |}} \rho   \scalebox{1.4}{\ensuremath{ \bm \rangle }}  
  = 
   \left[ \frac{1}{m} \hat{p} \hat{\lambda} - 
   V'(\hat{x}) \hat{\theta} \right] 
  \scalebox{1.3}{\ensuremath{ \bm |}} \rho \scalebox{1.4}{\ensuremath{ \bm \rangle }} 
,
\end{align}
where the position and momentum are given by the commuting operators $\hat{x}$ and $\hat{p}$ [Eq. (\ref{commutation-rels})].
In this limit the $x$-$p$ representation, $\Psi(x,p) =  \scalebox{1.4}{\ensuremath{ \bm \langle }}
 x p    \scalebox{1.4}{\ensuremath{ \bm | }}  \rho \scalebox{1.4}{\ensuremath{ \bm \rangle }}$, is the classical Koopman-von Neumann ``wave-function'' 
which is essentially the square root of the phase-space probability density. It has the differential equation  
\begin{align}
\label{Koopman-Eq-xp}
  \frac{\partial}{\partial t} 
 \scalebox{1.4}{\ensuremath{ \bm \langle }}
 x p    \scalebox{1.4}{\ensuremath{ \bm | }}  \rho 
 \scalebox{1.4}{\ensuremath{ \bm \rangle }}  = 
   \left[ - \frac{1}{m} p \frac{\partial }{\partial x} + 
   V'(x)  \frac{\partial}{ \partial p } 
  \right]  
 \scalebox{1.4}{\ensuremath{ \bm \langle }}
 x p  \scalebox{1.4}{\ensuremath{ \bm | }}  \rho 
 \scalebox{1.4}{\ensuremath{ \bm \rangle }},
\end{align}
Equation (\ref{Koopman-Eq-xp}) can be also obtained by taking the limit $\hbar \to 0$ of the Moyal equation (\ref{Moyal-Eq}) 
for the Wigner function $W(x,p)$. The corresponding positive phase-space probability 
density, $\rho(x,p) =  |\Psi(x,p)|^2 $, can be properly normalized  
\begin{align}
\label{rho-normalization}
\int \rho(x,p) dx dp = 1, 
\end{align}
and applying the chain rule to the definition of the density $\rho(x,p)$ one obtains the Liouville equation of 
classical mechanics, which strikingly is identical to that for the classical wave-function $\Psi(x,p)$~\cite{DaniloMauro2002}. 
Since Eq. (\ref{Koopman-Eq-xp}) is the equation obeyed by the classical probability density it is equivalent to an ensemble 
of Newtonian trajectories, as can be shown via the method of characteristics.
The classical evolution leaves the following cumulative function, time invariant 
\begin{align}
  C_{\rho}(\gamma,t) = \int_{\rho<\gamma} \rho(x,p,t)\, dx dp.  
\end{align}
This statement is proven by slicing the cumulative distribution for an arbitrarily 
small increment $\delta \gamma$
\begin{align}
   C_{\rho}(\gamma + \delta\gamma,t) -  C_{\rho}(\gamma,t) 
   = \int_{ \delta R }\!\!\!\! \rho(x,p,t)\, dx dp 
   \approx \gamma \!  \int_{ \delta R   } \!\!\!\! dx dp, 
\end{align}
where $\delta R$ is the region $\gamma < \rho <\gamma + \delta \gamma$. 
The latter integral measures the phase space volume where $\rho(x,p,t) \approx \gamma$, 
which is preserved according to Liouville's theorem, implying the time invariance of $C_{\rho}(\gamma,t)$.

The same arguments establish the time independence of the cumulative distribution
\begin{align}
  \label{KvN-CD}
  C_{\Psi}(\gamma,t) = \int_{\Psi<\gamma} \Psi(x,p,t)\, dx dp,  
\end{align}
for Koopman-von Neumann dynamics of real valued states $\Psi(x,p,t)$. Note 
that  $\Psi(x,p,t)$ is real for any time time if and only if the initial condition is real. Contrary 
to classical mechanics, quantum propagation of the Wigner function 
does not necessarily preserve the cumulative function. 
For example, a typical effect of quantum decoherence is the eventual elimination 
of any negativity in the Wigner function, 
\begin{align}
\label{negativity-eq}
  \mathcal{N}_{W}(t) = C_{W}(0,t)  = \int_{ W < 0 } W(x,p,t)\, dx dp.
\end{align}
Modern developments and applications of the Koopman-von Neumann classical mechanics
can be found in, e.g., Refs. \cite{Gozzi1988, Gozzi1989, Wilkie1997, Wilkie1997a, Gozzi2002, DaniloMauro2002, Deotto2003, Deotto2003a, Abrikosovjr2005, Blasone2005, Brumer2006, Carta2006, Gozzi2010, Gozzi2011, Cattaruzza2011, bondar2012operational,PhysRevA.88.052108}.

The Fokker-Planck equation of open classical dynamics can also be described in the present formalism 
\begin{align}
\label{Fokker-P}
 i {\partial_t} 
\rho(x,p)  = 
   \left[  \frac{1}{m} p  \hat{\lambda}  - 
   V'(x)  \hat{\theta} - i D \hat{\theta}^2
  \right]  \rho(x,p),
\end{align}
where $\hat{\lambda}$ and $\hat{\theta}$  are the differential operators defined in Eq. (\ref{phase-space-rep}). 
The classical limit of Eq. (\ref{B-D-eq}), governing quantum decoherence, recovers Eq. (\ref{Fokker-P})
as further discussed in Sec. \ref{SingleParticleSim}.  

\section{Spectral Split-Operator Methods}\label{SplitOpSec}

The unitary time-evolution operator, underlying the equation of motion (\ref{gen-motion}), for a time increment $dt$  is
\begin{align}
 U_{dt} = \exp \biggl( -i dt \left[ \frac{\hat{p}\hat{\lambda}}{m}   + \frac{V^{-} - V^{+}}{ \hbar } \right]  \biggr).
\end{align}
This operator can be approximated using the Trotter product~\cite{Trotter1959} in the limit of a small time 
increment either by the first-order scheme 
\begin{align}
 U_{dt} =  \exp \left( -i \frac{dt}{m} \hat{p}\hat{\lambda}  \right)
   \exp \left( - i \frac{dt}{ \hbar }( V^{-} - V^{+}  )   \right) + O(dt^2 ),
\end{align}
or by the second-order scheme~\cite{feit1982solution}
\begin{widetext}
\begin{align}
 U_{dt} = \exp \left( -i \frac{dt}{2m} \hat{p}\hat{\lambda}  \right)
   \exp \left( - i \frac{dt}{ \hbar }( V^{-} - V^{+}  )   \right)
   \exp \left( -i \frac{dt}{2m} \hat{p}\hat{\lambda}  \right)
  + O(dt^3 ).
\end{align}
Both factorizations are advantageous for numerical evaluations since the time-evolution propagator is expressed as a 
sequence of Fourier transforms $\mathcal{F}$ [see Eq. (\ref{Fourier-transforms})] and element-wise multiplications. Thus,
the first order scheme propagates the state in the $x$-$p$ representation according to 
\begin{align}
\label{B_plus}
 W(t + dt) = \mathcal{F}^{\lambda \to x} \exp \left( -i \frac{dt}{m} p \lambda  \right) \mathcal{F}_{\theta \to p}^{x \to \lambda}
   \exp \left( - i \frac{dt}{ \hbar }( V^{-} - V^{+}  )   \right)  \mathcal{F}_{p \to \theta}  W(t) , 
\end{align}
\end{widetext}
where $V^\pm =  V\left( x \pm \smallfrac{\hbar}{2} \theta  \right) $ have now become scalar functions, 
and $\mathcal{F}_{\theta \to p}^{x \to \lambda} = \mathcal{F}_{\theta \to p} \mathcal{F}^{x \to \lambda} =  \mathcal{F}^{x \to \lambda} \mathcal{F}_{\theta \to p}$ is a sequence of two Fourier transforms defined in Eq.(\ref{Fourier-transforms}). 
Numerical propagators for other representations of the Hilbert phase space can be developed in a similar fashion.

If the Wigner function $W(t)$ at a given point in time is stored in an array of length $N = N_p \times N_x$, then the total 
complexity of the propagator Eq. (\ref{B_plus}) is $O(N\log N)$ since it involves a sequence of two Fast Fourier 
Transforms \cite{frigo2005design} of $O(N\log N)$ complexity, and two element-wise multiplications of $O(N)$ complexity. 
The fast Fourier transform does not exactly coincide with the formal definition of the Fourier transform $\mathcal{F}$ because of the need to have one more element with negative  
frequency than with positive frequency. For convenience we thus give the propagators explicitly in terms of discrete position and momentum grids. We assume that both grids have an even number of points given respectively by $N_p$ and $N_x$, and denote the separation of the grid points by $\Delta x$ and $\Delta p$. In particular the grids are given by $x_n$ and $p_n$ with  
 \begin{align}
\Delta x = & 2 L_x /N_x  \qquad   \Delta p = 2 L_p /N_p \\
x_n = & -L_x + n \Delta x , \;\;\; n = 0, \ldots, N_x-1, \\
p_m = & -L_p+ m \Delta p, \;\;\; m = 0, \ldots, N_p-1, 
\end{align}
where $L_x$ and $L_p$ define the window of interest in the phase space.
The Wigner function is actually stored with the grid elements in a different order, in that the negative grid points are stored in 
the second half of the grid. This order is given by $W_{kj} = W(\tilde{x}_j,\tilde{p}_k)$ with 
\begin{align}
\label{no-fftshift}
  \tilde{x}_j & =  \left\{ 
\begin{array}{ll}
  x_{j+N_x/2}  &   \mbox{for}  \;\;\;  j = 0, \ldots, \smallfrac{N_x}{2} - 1    \\   
  x_{j-N_x/2}   &   \mbox{for}  \;\;\;  j = \smallfrac{N_x}{2}, \ldots, N_x - 1 
\end{array} 
\right.
\end{align}
and with the corresponding relationship between $\tilde{p}_k$ and $p_m$. Note that the Wigner function at the origin of the coordinate 
system, $W(0,0)$, is now stored at the edge of the grid as $W_{00}$. The reason for this new grid ordering is that it is the 
natural ordering upon which to apply the fast Fourier transform. It is, of course, not the natural ordering to use 
in displaying the Wigner function, so we transform from the $j,k$ ordering to the $n,m$ ordering before plotting. 
This transformation is called an ``FFT shift'' and is characterized for being a self-inverse function. 
It is usually provided in libraries that implement the fast Fourier transform. However it should be noted that some 
implementations of the  ``FFT shift''  store an extra copy of  the Wigner function, which can be prohibitively 
expensive, which is why the user may need to make explicit use of Eq. (\ref{no-fftshift}). We provide 
a Python implementation of the unitary propagation for a single-particle in the Appendix \ref{source-code}.    

In the case of other representations, e.g., $x-\theta$, the grid discretization step size $\Delta\theta$ is given by 
\begin{align}
 \Delta\theta  =  2 \pi / L_p ,  \qquad L_{\theta} = \Delta\theta  N_p/2;
\end{align}
whereas in the $\lambda-p$ representation
\begin{align}
  \Delta\lambda =  2 \pi / L_x ,  \qquad L_{\lambda} = \Delta\lambda N_x/2.
\end{align}
If the system's initial condition is given by a wave function known analytically, 
then $B(x,\theta)$ can be readily constructed by Eq. (\ref{B-def}), whereas the calculation of 
the corresponding Wigner distribution requires an additional Fourier transform (\ref{B-to-W}).

\subsection{Solving master equations}

The split-operator method presented above can be extended to handle non-unitary open quantum system dynamics. For 
example, the first-order split-operator method for evolving the master equation given 
in Eq. (\ref{ExtendedLindbladDecoherence}) is 
\begin{widetext}
\begin{align}
\label{DecoherenceSplitOperator}
	W(t+ dt) = 
		 \mathcal{F}^{ \lambda \to x}   \exp{ \left( -\frac{i dt}{m} p \lambda  \right)} 
	        \mathcal{F}^{x \to\lambda }_{ \theta \to p}  \exp{ \left( -\frac{i dt}{\hbar}\left[  V^{-}   - V^+  \right]
  - dt D \theta^2    \right)} \mathcal{F}_{p \to \theta}  W(t).
\end{align}
Similar techniques can be used for solving the classical Liouville 
equation \cite{PhysRevE.51.821, dattoli1996symmetric, gomez2014split}, and can be extended to the  Koopman-von Neumann 
equation (\ref{Koopman-Eq-xp}). However, Liouville-like equations can only be solved exactly for a finite time 
on a fixed grid, due to the development of increasingly fine structure, 
know as {\it velocity filamentation}~\cite{kellogg1965some}. This issue can be handled by filtering the phase-space 
distribution so as to remove high-frequency (spatial) structure. In the $x$-$\theta$ representation this results 
in the following propagation scheme for $\rho(t)$ and $\Psi(t)$~\cite{Klimas1987202, narayan2009deterministic} 
\begin{align}
 \label{KvN-propagator}
	\left\{ \rho(t+dt) \atop \Psi(t+dt) \right\} = 
		\mathcal{F}^{\lambda \to x}  \exp{ \left( -\frac{i dt}{m} p \lambda  \right)}
                  \mathcal{F}^{x \to\lambda }_{ \theta \to p}  
	         \exp{ \left( - i dt V'(x) - \delta D \theta^2 \right)}    \mathcal{F}_{p \to \theta}  
              	\left\{ \rho(t) \atop \Psi(t) \right\}   ,
\end{align}
\end{widetext}
valid for both the  classical probability density $\rho(x,p)$ and the Koopman-von Neumann wave function $\Psi(x,p)$. 
This propagator is equivalent to the evolution of the Fokker-Planck equation (\ref{Fokker-P}); the diffusion term 
in the Fokker-Planck equation washes out the fine structure. 
A similar numerical trick is used to develop efficient numerical methods for the Hamiltonian-Jacobi 
equation \cite{Sethian1999}. 

The Caldeira-Legget master equation, Eq. (\ref{D-Caldeira-Legget}), can be implemented by separating 
the effects of decoherence and dissipation. The second term in Eq. (\ref{D-Caldeira-Legget}), generating decoherence, 
has already been treated in Eq. (\ref{DecoherenceSplitOperator}). The first term in Eq. (\ref{D-Caldeira-Legget}) 
describes energy exchange with the bath, and could be evaluated by specially designed finite difference 
schemes \cite{vesely1994computational, Collins2014299, grossmann2009finite}, although these require large 
grid sizes to achieve numerical stability.

To overcome this limitation we now present a stable method enabling, for the first time, higher dimensional 
simulations (see Sec.\ref{TwoParticleSim}). The time evolution induced by a general dissipator operator $\hat{\mathcal{C}}$ is
\begin{align}
   \scalebox{1.4}{\ensuremath{ \bm | }} \rho(t+dt )  \scalebox{1.4}{\ensuremath{ \bm \rangle }}
   & = e^{ dt \hat{\mathcal{C}} } 
  \scalebox{1.4}{\ensuremath{ \bm | }} \rho(t)  \scalebox{1.4}{\ensuremath{ \bm \rangle }} \nonumber \\ 
  & =  \left(
 1 + dt\, \hat{\mathcal{C}} [ 1 + dt \, \hat{\mathcal{C}}/2  ] 
 \right)
   \scalebox{1.4}{\ensuremath{ \bm | }} \rho(t)  \scalebox{1.4}{\ensuremath{ \bm \rangle }}
 + O(dt^3). \label{gendisprop}
\end{align}
For the energy exchange term in the Caldeira-Legget model, Eq. (\ref{D-Caldeira-Legget}), 
we have $ \hat{\mathcal{C}} =  2i \gamma \hat{\theta} \hat{p}$. Using Eq. (\ref{gendisprop}) we can propagate the Wigner function in two steps as 
\begin{align}
\label{second-order-C}
 W(t+dt) & = W(t) + 2i dt \gamma\, \hat{\theta}\hat{p}\, W^{(1)}(t) , \\
  W^{(1)}(t) & =  W(t) + i dt \gamma\, \hat{\theta}\hat{p}\, W(t)
\end{align}
where a sequence of $\theta \to p$ Fourier transforms is used to calculate the required 
operator product:  
\begin{align}
  \hat{\theta} \hat{p}  W(t) 
 =  \mathcal{F}_{\theta \to p }\,    \theta\,  \mathcal{F}_{p \to \theta }\,   p W(t). 
\end{align}
We note that the second-order scheme (\ref{second-order-C}) is sufficient for the simulations 
in Sec. \ref{SingleParticleSim} and \ref{TwoParticleSim}; nevertheless, 
higher order corrections can be recursively included if needed.  

\section{Single-particle Systems}
\label{SingleParticleSim}

\begin{figure*}
  \includegraphics[width=1.\hsize]{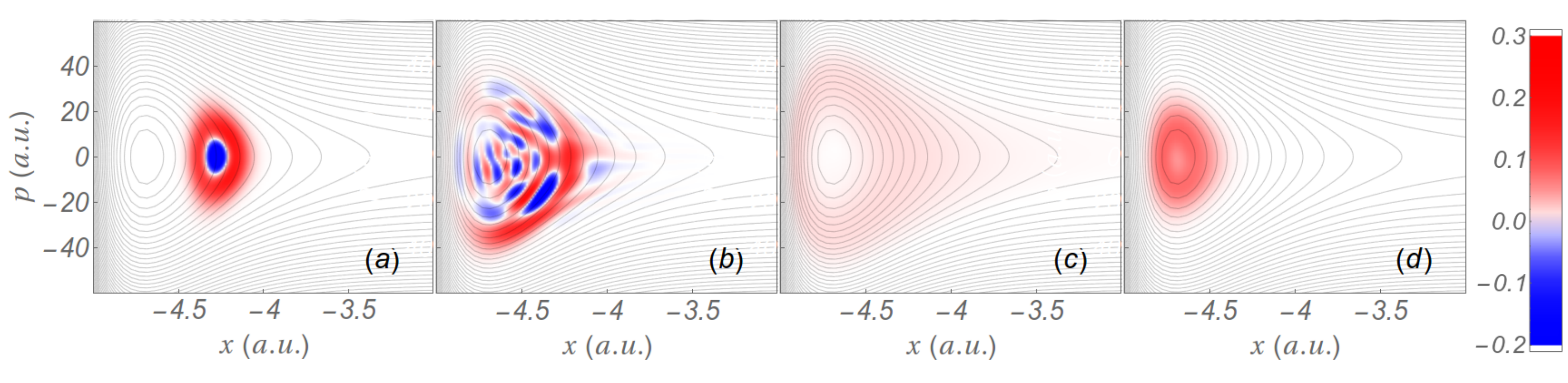}
  \caption{(Color online) Various quantum dynamics in the Morse potential $V(x)$ given in Eq.(\ref{morse-pot}).
  The contour lines represent level sets of the classical energy $H(x,p) = p^2 / (2m) + V(x)$.  
 (a)  The initial Wigner function (WF) at $t=0~\mbox{a.u.}$
 (b)  The WF at time $t = 40,400~\mbox{a.u.}$ after unitary evolution employing Eq.(\ref{B_plus}). 
 (c)  The WF at $t = 40,400~\mbox{a.u.}$ after unitary evolution with additional decoherence in the position basis. The diffusion coefficient is $D= 2.70 \times 10^{-3}~\mbox{a.u.}$ and we use  
     the propagator in Eq.(\ref{DecoherenceSplitOperator}).
 (d) The WF at time $t=40,400~\mbox{a.u.}$ after unitary evolution with energy damping given by the Caldeira-Legget model with temperature $T = 300 K$, diffusion $D= 2.70 \times 10^{-3}~\mbox{a.u.} $ and inverse damping coefficient $\gamma^{-1}  = 41,341~\mbox{a.u.} = 1~\mbox{ps}$. 
All these simulations were performed with a grid of $512 \times 1024 $.}
\label{fig:figure_morse1}
\end{figure*}

In this section, we apply the numerical methods developed in Sec.~\ref{SplitOpSec} to propagate a single-particle under 
various interactions with the environment. We consider the model for vibrational diatomic molecular dynamics: a 
particle with mass $ m = 58752~\mbox{a.u.} $  (we use atomic units (a.u.) throughout) moving in a Morse potential given by 
\begin{align} 
\label{morse-pot}
V(x) =  V_0[\exp(-2a [x-r_e] ) -2 \exp(a[x-r_e])  ], 
\end{align}
with  $V_0 = 0.6~\mbox{eV} = 0.0220~\mbox{a.u.} $, $a = 2.5~\mbox{a.u.}$ and $r_e = -4.7~\mbox{a.u.}$ The Wigner function for the initial state 
is shown in Fig. \ref{fig:figure_morse1}(a). This initial state corresponds to the first-exited state 
of the Morse potential displaced by $x_0 = 4.3~\mbox{a.u.} $, and is given by  
\begin{align}
\psi_1(x)  =  N z^{L-n-1/2}e^{-z/2} \left( 1 - \frac{L\exp(-a[x-x_0])}{L-1}   \right) , 
\end{align}
where $L= \sqrt{2 m V_0}/a$ and $N$ is a normalization constant. This state possesses significant 
negativity, defined in Eq. (\ref{negativity-eq}), and we propagate it according to three different dynamical 
equations: (i) unitary evolution, Eq. (\ref{gen-motion}), resulting in the final state 
shown in Fig. \ref{fig:figure_morse1} (b); (ii) decoherent dynamics given by Eq. (\ref{B-D-eq}), 
resulting in the final state shown in Fig. \ref{fig:figure_morse1}(c); (iii) Evolution under 
the Caldeira-Legget master equation, Eq. (\ref{Caldeira-Legget-eq}), resulting in the final state 
given in Fig. \ref{fig:figure_morse1}(d). 

These simulations provide an opportunity to observe the emergence of the classical world as a result of the interactions 
with the 
environment~\cite{Bhattacharya00, PhysRevLett.88.040402, Bhattacharya03, RevModPhys.75.715, Everitt09, jacobs2014quantum}. 
In particular they illustrate how decoherence eliminates the negative regions of the Wigner function. The final state under 
purely unitary evolution in Fig. \ref{fig:figure_morse1}(b) contains significant negativity  (\ref{negativity-eq}), 
while the states in the presence of interactions with the environment evolve to entirely 
positive states as seen in Figs  \ref{fig:figure_morse1}(c) and  \ref{fig:figure_morse1}(d). 

\begin{figure*}
  \includegraphics[width=0.9\hsize]{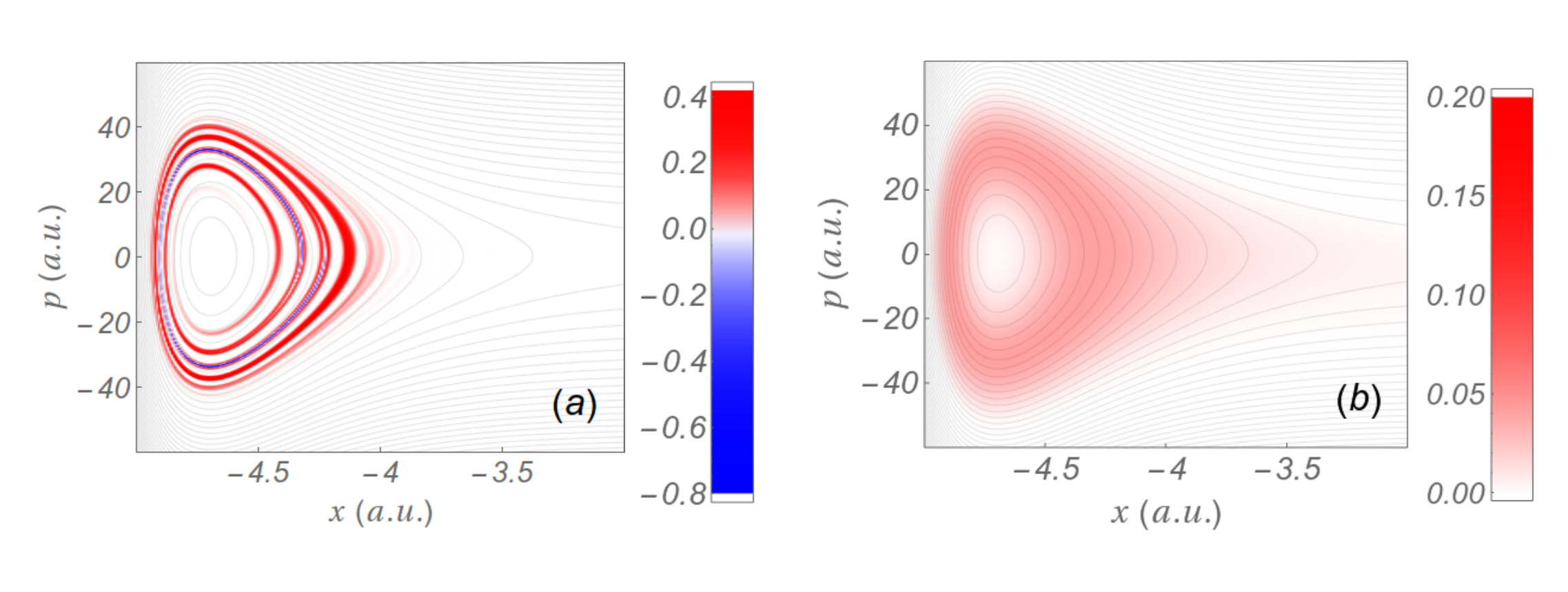}
  \caption{ (Color online)   
   (a)  
    Koopman-von Neuman state   classically propagated at  $t=40400~\mbox{a.u.}$,   with regularization   
    diffusion coefficient $\delta D = 1.5\times 10^{-6}~\mbox{a.u.} $ in a grid $768 \times 6144 $.
   (b) Corresponding classical state propagated according to the Fokker-Planck equation 
   with decoherence (diffusion) coefficient equal to $D = 2.61 \times 10^{-3}~\mbox{a.u.} $ in a grid $512 \times 1024$.
 }
 \label{fig:figure_morse2}
\end{figure*}

We also propagated the initial state shown in Fig.~\ref{fig:figure_morse1}(a) using   
i) the classical Koopman-von Neumann evolution, Eq. (\ref{Koopman-Eq-xp}), regularized to handle the velocity filamentation 
(see the discussion in Sec.~\ref{SplitOpSec} ), and ii) the Fokker-Planck evolution (\ref{Fokker-P}) with the same diffusion 
coefficient as used for the open-system evolution shown Fig.~\ref{fig:figure_morse1}(c). The result of the Koopman-von Neumann 
evolution is shown in Fig.~\ref{fig:figure_morse2}(a) and that of the Fokker-Planck equation 
is shown in Fig.~\ref{fig:figure_morse2}(b). 
A comparison of the final states in Fig.~\ref{fig:figure_morse1} and Fig.~\ref{fig:figure_morse2} shows that a quantum 
state undergoing decoherence converges to the solution of the Fokker-Planck equation, rather than to the corresponding 
Koopman-von Neumann state. The reason for this is that the decoherence is a measurement process and induces quantum 
back-action noise that is equivalent to diffusion, and the Fokker-Planck equation correctly includes this diffusion. The 
classical limit is defined as that in which the action of a system is sufficiently large that the decoherence needed to 
transform the motion into classical dynamics induces diffusion that is negligible in comparison. In that case the open-system 
dynamics converges to the Koopman-von Neumann evolution (equivalently the classical Liouville evolution) because the effect 
off the diffusion is negligible. We note that the color scales in Figs.~\ref{fig:figure_morse1} 
and \ref{fig:figure_morse2} differ due to the different normalization conventions for the Wigner 
function  (\ref{W-normalization}) and the Koopman-von Neumann state (\ref{rho-normalization}). 

While the quantum evolution has a bound on the smallest structure in the phase space \cite{zurek2001sub}, 
the Koopman-von Neumann evolution develops an ever finer structure,
even for a non-chaotic classical system (see Fig. \ref{fig:figure_morse2}(b)). As a result the Koopman-von Neumann 
simulations required significantly larger grids than either the quantum or Fokker-Planck simulations.
 
\begin{figure}
  \includegraphics[width=0.7\hsize]{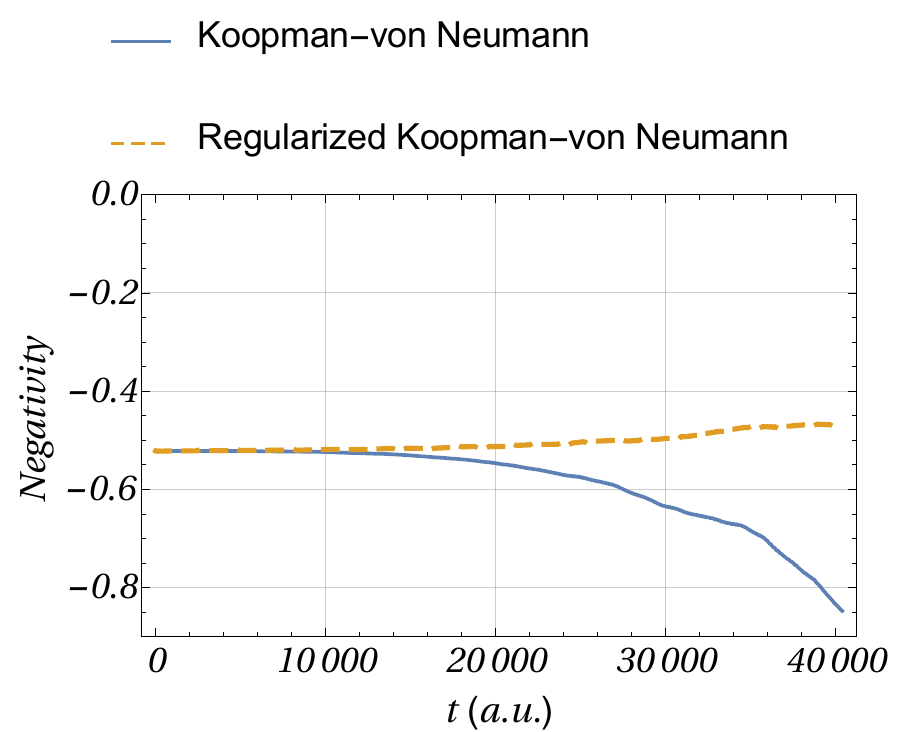}
\caption{ Negativity as a function of time for 
i) the regularized Koopman-von Neumann propagation with decoherence coefficient $\delta D = 1.5\times 10^{-6}~\mbox{a.u.}$ 
(dashed line), ii) Koopman-von Neumann propagation without regularization (solid line),
The regularized Koopman-von Neumann propagator maintains an approximately constant negativity, contrary to 
the monotonic increase given by the un-regularized version. }
\label{fig:Negativity-area}
\end{figure}

The need to regularize the Koopman-von Neumann propagator, Eq. (\ref{KvN-propagator}), is illustrated 
in Fig.~\ref{fig:Negativity-area}, where we can see that without regularization the propagator fails to maintain 
the negativity (Eq. (\ref{negativity-eq})), 
while the regularized version, in which a small amount of decoherence is added, keeps the negative area approximately 
constant for long times. In addition, 
Fig.~\ref{fig:Negativity-area} shows that a larger decoherence rate quickly eliminates all the negativity.  

\section{ Two-particle Systems}
\label{TwoParticleSim}

A two-particle quantum system in phase space involves four degrees of freedom (i.e., $x$, $p_x$, $y$, and $p_y$), and has rarely 
been simulated even for closed system dynamics \cite{filinov1995quantum, dittrich2010semiclassical}. Here we 
study open system dynamics within 
the Caldeira-Legget master equation, which has never been attempted, to the best of our knowledge. Even so, we are able to run 
these simulations on a typical desktop machine. To do this an efficient  use of memory becomes critical, and because of this we 
perform the computations employing single precision arithmetic (32 bit floats). We use a grid which is $128\times 192 \times 128 \times 192$ and 
occupies 4.7GB of memory. Two copies of the state are needed according to Eq. (\ref{second-order-C}). The resulting simulation 
of the Caldeira-Legget evolution remains numerically stable even for the time increment $dt = 0.01 a.u.$, which is unattainable 
by alternative methods~\cite{vesely1994computational, Collins2014299, grossmann2009finite}. 
 
The two particle Wigner function, $W(x,p_x,y,p_y)$, expressed through the density matrix 
\begin{widetext}
\begin{align}
 W_2(x,p_x,y,p_y) 
 =  \frac{1}{(2\pi)^2} \int   \langle  x - \frac{\hbar}{2}\theta_x ,   y - \frac{\hbar}{2}\theta_y  
  | \rho  | x + \frac{\hbar}{2}\theta_x ,   y + \frac{\hbar}{2}\theta_y   \rangle \,  e^{i p_x \theta_x  + i p_y \theta_y  } d \theta_x d\theta_y ,
\end{align}
\end{widetext}
can be reduced to the following single particle Wigner functions, 
\begin{equation}
   W_x(x,p_x) = \int W_2 \, dy dp_y , \quad W_y(y,p_y) = \int W_2 \, dx dp_x  , 
\end{equation}
which are more easily visualized. Note that even if the two particle state is pure the reduced states may be mixed. The purity of an arbitrary state in the phase space is given by 
\begin{align}
 \mathcal{P} = 2 \pi \int W^2(x,p) dx dp,
\end{align}
where the maximum value  $\mathcal{P} = 1 $ is attained for pure states only.

\begin{figure*}
  \centering
      \includegraphics[width=1\hsize]{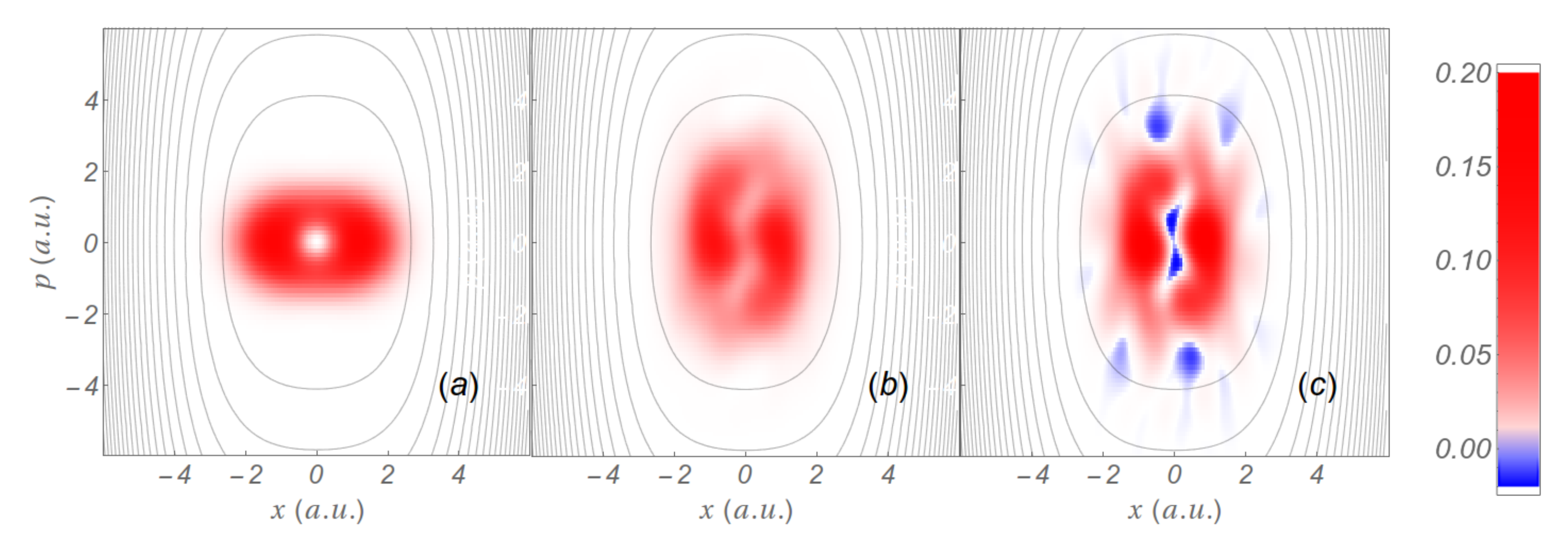}
  \caption{
(a) Initial reduced fermionic-like state for both particles ($W_x = W_y$). (b) Reduced state $W_x$ at $t=5.0~\mbox{a.u.}$. (c) Reduced state $W_y$ at  $t=5.0~\mbox{a.u.}$. 
}
  \label{fig:fermions}
\end{figure*}

\begin{figure}
  \centering
      \includegraphics[width=0.7\hsize]{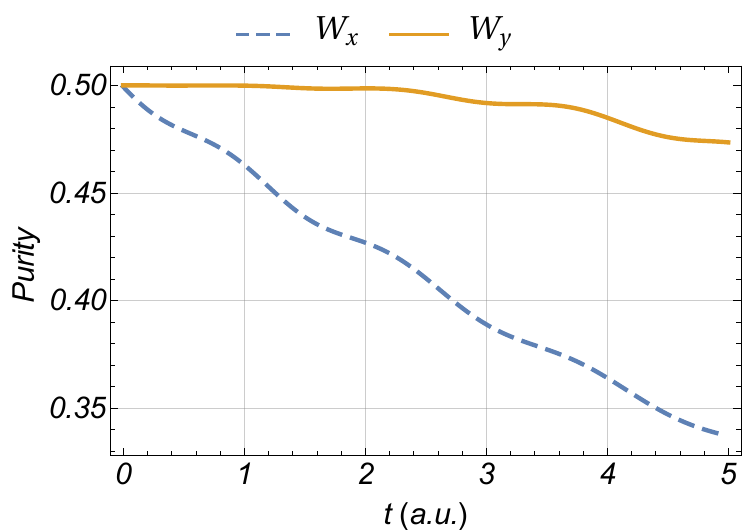}
  \caption{Evolution of the purity for the Fermionic-like reduced states $W_x$ (dashed line) and $W_y$ (solid line).}
  \label{fig:FermionPurity}
\end{figure}

Here we simulate a two particle system evolving in the anharmonic potential 
\begin{align}
 V(x,y) = \frac{1}{2} \left( x^2 + y^2 \right) + \frac{1}{10}\left( x^4  + y^4 + x y \right), 
\end{align}
where the first particle interacts with an environment and as result is subject to the Caldeira-Leggett master 
equation, Eq. (\ref{Caldeira-Legget-eq}). The Caldeira-Leggett dynamics is similar to a position 
measurement because it decoheres the system in the position basis. 
We chose $D = 0.04\,~\mbox{a.u.}$ and $\gamma = 1./12.5\,~\mbox{a.u.} $. The second particle does not interact with 
the environment, and is only affected by the latter through its interaction with the first particle. Such coupled systems play an important role in describing quantum measurements~\cite{wheeler1983quantum, jacobs2006straightforward, jacobs2014quantum, clerk2010introduction, bose1999scheme}. 
The initial state is chosen to be an antisymmetric pure entangled state [Figs. \ref{fig:fermions}(a)]
\begin{align}
 \psi_F(x,y) =& \frac{1}{\sqrt{2}}
\left[ \psi_1(x)  \psi_2(y)  -  \psi_1(y )  \psi_2(x) \right], 
\end{align}
where $\psi_1(x)$ is a Gaussian centered at $x=1$, and $\psi_2(x)$ is another Gaussian centered at
 $y=-1$. Both reduced single particle Wigner functions are identical for this state. 
However, due to the environment interaction with the first particle, the reduced Wigner functions $W_x$ and $W_y$ are not 
equal at later times, and this is shown in Fig.~\ref{fig:fermions}(b) and (c). Moreover,  $W_y$ has a larger negativity 
than $W_x$, indicating that it preserves more of its initial quantum nature. Figure~\ref{fig:FermionPurity} shows 
how the purity of both reduced states evolves with time.  

\section{Conclusion}
\label{conc}

We have presented a flexible and powerful numerical toolbox for simulating open quantum systems in terms of the Wigner 
function. These methods significantly reduce the numerical resources required for exact simulation of open systems in phase 
space, and the method we have presented for solving the Caldeira-Leggett master equation enjoys higher stability than currently 
available methods. Illustrative examples were provided for single- and two-particle systems that can be evaluated on a typical 
desktop computer. In these examples we illustrated the emergence of a positive Wigner function as a result of decoherence and 
compared it with the classical Koopman-von Neumann and Fokker-Planck evolutions. These simulations confirm that quantum 
evolution with decoherence approaches classical Fokker-Planck dynamics. 

\emph{Acknowledgments}. The authors acknowledge financial support 
from (HR) NSF CHE 1058644, (RC) DOE DE-FG02-02-ER-15344 and (DB) ARO-MURI W911-NF-11-1-2068. 


%

\newpage

\appendix

\section{Appendix: Python code for a single particle}

\label{source-code}
\begin{lstlisting}[language=Python]

# coding: utf-8
## Wigner Propagator
# In[1]:

import matplotlib.pyplot as plt
from scipy.special import laguerre
import numpy as np
import scipy.fftpack as fftpack
from scipy.special import hyp1f1

# In[2]:

def Fourier_XTheta_to_XP(M):
    return fftpack.ifft( M , axis=0 )

# In[3]:

def MorsePsi( n , morse_a, morse_Depth, mass, r0, r):
    x = (r-r0)*morse_a
    LAMBDA = np.sqrt(2*mass*morse_Depth )/morse_a
    z = 2.*LAMBDA*np.exp(-x)
    return z**( LAMBDA - n - 0.5 )*np.exp( -0.5*z )*hyp1f1( -n , 2*LAMBDA - 2*n  , z )

# In[4]:

def Morse_Wigner(mass, morse_a, morse_Depth, X, Theta):
        x_init = -4.3
        p_init =  0.
        n = 1    
        
        psiLeft = MorsePsi(n, morse_a, morse_Depth, mass, x_init, X - 0.5*Theta)
        
        psiRight= MorsePsi(n, morse_a, morse_Depth, mass, x_init, X + 0.5*Theta)
        
        B = psiLeft*psiRight.conj() 
        
        W = Fourier_XTheta_to_XP( B )
        
        #norm = np.sum( W )*dX*dP
        #W /= norm
        
        return W


# In[5]:

X_gridDIM = 2*512     # Discretization grid size in X
P_gridDIM =   512     # Discretization grid size in P
        
X_amplitude  = 6      # Window range -X_amplitude to X_amplitude
P_amplitude  = 80     # Window range -P_amplitude to P_amplitude
        
dt= 1.01
        
timeSteps =    1000   # Number of iterations
        
mass = 32.*1836       # Mass of the particle

# Decoherence parameter

timeDamping  = 10.**(-12) * 1./( 2.418884326505*10.**(-17) )         # 1 ps
temperatureT = 300. * 1.3806488*10**(-23) /( 4.35974417*10**(-18)  )
gammaDamping = 1./timeDamping  
D_Theta  = 2*mass*temperatureT*gammaDamping   #  Decoherence parameter  
        
print ' gamma = ', gammaDamping , ' D_Theta = ', D_Theta #, ' D_Lambda = ', D_Lambda
print ' final time = ', timeSteps*dt ,'a.u.   =', timeSteps*dt*2.418884326505*10.**(-17),' s'
        
# Potential parameters

morse_a = 2.5         
morse_Depth = 0.6/27.211   # 0.4 eV
morse_x = -4.7


# In[6]:

def truncationFunction(x):
    return 1000.*np.arctan( x/1000. )

def Potential(x):
    return truncationFunction(
            morse_Depth*(np.exp(-2*morse_a*(x-morse_x)) - 2.*np.exp(-morse_a*(x-morse_x  )  )   )
            )


# In[7]:

# Discretization resolution

dX =  2.*X_amplitude/float(X_gridDIM)
dP =  2.*P_amplitude/float(P_gridDIM)

dTheta  = 2.*np.pi/(2.*P_amplitude)
Theta_amplitude = dTheta*P_gridDIM/2.
dLambda = 2.*np.pi/(2.*X_amplitude)
Lambda_amplitude = dLambda*X_gridDIM/2.

# vectors with range of coordinates
X_range      =  np.linspace(-X_amplitude      , X_amplitude  -dX , X_gridDIM )
Lambda_range =  np.linspace(-Lambda_amplitude , Lambda_amplitude-dLambda  , X_gridDIM)

Theta_range  = np.linspace(-Theta_amplitude  , Theta_amplitude - dTheta , P_gridDIM)
P_range      = np.linspace(-P_amplitude      , P_amplitude-dP           , P_gridDIM)

# matrices of grid of coordinates
X      = fftpack.fftshift(X_range)[np.newaxis,:]
Theta  = fftpack.fftshift(Theta_range)[:,np.newaxis]

Lambda = fftpack.fftshift(Lambda_range)[np.newaxis,:]
P      = fftpack.fftshift(P_range)[:,np.newaxis]


# In[8]:

#  Center of the wave packet
x_init = -4.3
p_init =  0.

# Standard deviation in x

mass  = 58762

# initialization wigner function

W_init = Morse_Wigner( mass, morse_a, morse_Depth, X, Theta)

norm   = np.sum( W_init )*dX*dP
W_init /= norm


# In[9]:

# Ploting potential 
fig, ax = plt.subplots(figsize=(10, 5))
ax.plot( X_range,   Potential(X_range) )
ax.set_xlim(-5,4)
ax.set_ylim(-0.025,0.001)
ax.set_xlabel('x')
ax.set_ylabel('V')
ax.grid('on')
ax.set_aspect(200.)


# In[10]:

# Matrix grid for the classical Hamiltonian

Hamiltonian_Classical = fftpack.fftshift( P**2/(2*mass) + Potential(X) )


# In[11]:

# Iteration

timeRangeIndex = range(0, timeSteps+1)

print ' ------------------------------------------------------------------------------- '
print '     Split Operator Propagator                                                   '
print ' ------------------------------------------------------------------------------- '

W = W_init

expPLambda = np.exp(-1j*dt*P*Lambda/mass)
expPotential = np.exp( -dt*1j*(Potential(X-Theta/2.) - Potential(X+Theta/2.)) )

for tIndex in timeRangeIndex:
    if tIndex%100==0:
        print '  tIndex  = ', tIndex
    t = (tIndex)*dt
    # p x  ->  p lambda 
    W = fftpack.fft( W, axis = 1 )
    W *=expPLambda

    # p lambda  ->  p x
    W = fftpack.ifft( W, axis = 1 )
    
    # p x -> theta x
    W = fftpack.fft( W, axis = 0 )
    W *= expPotential

    # theta x  ->  p x
    W = fftpack.ifft( W, axis = 0 )

    # Normalization
    norm = np.sum( W )*dX*dP
    W /= norm

W_end = np.real(W)        


### Plotting

# In[12]:

def PlotWigner(W):
    
    W = fftpack.fftshift( W )
    
    global_color_max = 0.17          #  Maximum value used to select the color range
    global_color_min = -0.31         # 
        
    print 'min = ', np.min( W ), ' max = ', np.max( W )
    
    print 'normalization = ', np.sum( W_end )*dX*dP
    zero_position =  abs( global_color_min) / (abs( global_color_max) + abs(global_color_min)) 
    wigner_cdict = {'red' 	: 	((0., 0., 0.),
							(zero_position, 1., 1.), 
							(1., 1., 1.)),
					'green' :	((0., 0., 0.),
							(zero_position, 1., 1.),
							(1., 0., 0.)),
					'blue'	:	((0., 1., 1.),
							(zero_position, 1., 1.),
							(1., 0., 0.)) }
    wigner_cmap = plt.cm.colors.LinearSegmentedColormap('wigner_colormap', wigner_cdict, 256)

    fig, ax = plt.subplots(figsize=(12, 5))

    #wigner_cmap = ax.colors.LinearSegmentedColormap('wigner_colormap', wigner_cdict, 256)
    
    x_min = -X_amplitude
    x_max = X_amplitude - dX
    
    p_min = -P_amplitude
    p_max = P_amplitude - dP
    
    cax = ax.imshow( W ,origin='lower',interpolation='none',
                    extent=[ x_min , x_max, p_min, p_max],
                    vmin= global_color_min, vmax=global_color_max, cmap=wigner_cmap)
    
    cbar = fig.colorbar(cax, ticks=[-0.3, -0.2,-0.1, 0, 0.1, 0.2 , 0.3])
    ax.set_aspect(1./64.)
    ax.set_xlim( -5.3, -3)
    
    ax.contour( Hamiltonian_Classical ,
    np.arange(-1, 1, 0.01 ),origin='lower',extent=[x_min,x_max,p_min,p_max],
    linewidths=0.25,colors='k')
    
    return fig


# In[13]:

init_fig = PlotWigner( W_init.real );
init_fig.savefig("Morse_init.png");

# In[14]:

end_fig = PlotWigner( W_end.real );

end_fig.savefig("Morse_end.png");

# In[14]:



\end{lstlisting}


\end{document}